\documentclass[final,5p,times,twocolumn,sort&compress]{elsarticle}
\usepackage{setspace}

\usepackage{amssymb}
\usepackage[version=3]{mhchem}
 \usepackage[utf8]{inputenc}
\usepackage{graphicx}
\usepackage{dcolumn}
\usepackage{float}
\usepackage{mathptmx, courier, pifont}
\usepackage[scaled=0.92]{helvet}
\usepackage[T1]{fontenc}
\usepackage{textcomp}
\usepackage{color}
\usepackage[normalem]{ulem}
\usepackage{booktabs,longtable}
\usepackage{rotating}
\usepackage{multirow}
\usepackage{datetime}

\usepackage[dvipsnames]{xcolor}
\usepackage[colorlinks=true,urlcolor=Blue,linkcolor=Blue]{hyperref}
\usepackage[all]{hypcap}

\newcommand{\red}{}
\newcommand{\redx}{}

\journal{Physics Letters B}

\begin{document}

\begin{frontmatter}


\title{\red{New Test of Modulated} Electron Capture Decay of Hydrogen-Like $\ce{^{142}Pm}$ Ions\red{: Precision Measurement of Purely Exponential Decay}}
\author[rvt]{F.~C.~Ozturk\corref{cor1}}
\ead{fcozturk@istanbul.edu.tr}
\author[rvt]{B.~Akkus}
\author[cern]{D.~Atanasov}
\author[focal]{H.~Beyer}
\author[]{\framebox{F.~Bosch}}
\author[jlu2,saclay]{D.~Boutin}
\author[focal,jlu1]{C.~Brandau}
\author[austria]{P.~B{\"u}hler}
\author[rvt]{R.~B.~Cakirli}
\author[focal,impcas]{R.~J.~Chen}
\author[impcas,fffff]{W.~D.~Chen}
\author[focal,impcas]{X.~C.~Chen}
\author[triumf]{I.~Dillmann}
\author[focal]{C.~Dimopoulou}
\author[focal]{W.~Enders}
\author[focal]{H.~G.~Essel}
\author[d]{T.~Faestermann}
\author[e]{O.~Forstner}
\author[focal,impcas]{B.~S.~Gao}
\author[focal]{H.~Geissel}
\author[d]{R.~Gernh{\"a}user}
\author[focal,guf]{R.~E.~Grisenti}
\author[focal]{A.~Gumberidze}
\author[focal,guf]{S.~Hagmann}
\author[guf]{T.~Heftrich}
\author[focal]{M.~Heil}
\author[e]{M.~O.~Herdrich}
\author[focal]{P.-M.~Hillenbrand}
\author[ccrf]{T.~Izumikawa}
\author[]{\framebox{P.~Kienle}}
\author[austria]{C.~Klaushofer}
\author[focal]{C.~Kleffner}
\author[focal]{C.~Kozhuharov}
\author[focal,jlu2]{R.~K.~Kn{\"o}bel}
\author[focal]{O.~Kovalenko}
\author[cern]{S.~Kreim}
\author[focal]{T.~K{\"u}hl}
\author[edinb]{C.~Lederer-Woods}
\author[focal]{M.~Lestinsky}
\author[focal]{S.~A.~Litvinov}
\author[focal]{Yu.~A.~Litvinov\corref{cor1}}
\ead{Y.litvinov@gsi.de}
\author[impcas]{Z.~Liu}
\author[impcas]{X.~W.~Ma}
\author[d]{L.~Maier}
\author[guf]{B.~Mei}
\author[saitama]{H.~Miura}
\author[focal]{I.~Mukha}
\author[d]{A.~Najafi}
\author[tsukuba]{D.~Nagae}
\author[saitama]{T.~Nishimura}
\author[focal]{C.~Nociforo}
\author[focal]{F.~Nolden}
\author[niigata]{T.~Ohtsubo}
\author[rvt]{Y.~Oktem}
\author[saitama]{S.~Omika}
\author[tsukuba]{A.~Ozawa}
\author[focal]{N.~Petridis}
\author[kracow]{J.~Piotrowski}
\author[guf]{R.~Reifarth}
\author[focal]{J.~Rossbach}
\author[focal]{R.~S\'anchez}
\author[focal]{M.~S.~Sanjari}
\author[focal]{C.~Scheidenberger}
\author[focal]{R.~S.~Sidhu}
\author[focal]{H.~Simon}
\author[focal]{U.~Spillmann}
\author[focal]{M.~Steck}
\author[focal,e,f]{Th.~St{\"o}hlker}
\author[beihang]{B.~H.~Sun}
\author[rvt]{L.~A.~Susam}
\author[saitama,riken]{F.~Suzaki}
\author[saitama]{T.~Suzuki}
\author[spbu]{S.~Yu.~Torilov}
\author[focal,jlu1]{C.~Trageser}
\author[paris]{M.~Trassinelli}
\author[focal,e]{S.~Trotsenko}
\author[focal,impcas]{X.~L.~Tu}
\author[surrey]{P.~M.~Walker}
\author[impcas]{M.~Wang}
\author[focal,e]{G.~Weber}
\author[focal]{H.~Weick}
\author[focal]{N.~Winckler}
\author[focal]{D.~F.~A.~Winters}
\author[edinb]{P.~J.~Woods}
\author[saitama]{T. Yamaguchi}
\author[impcas]{X.~D.~Xu}
\author[impcas]{X.~L.~Yan}
\author[impcas]{J.~C.~Yang}
\author[impcas]{Y.~J.~Yuan}
\author[impcas,emmi]{Y.~H.~Zhang}
\author[impcas]{X.~H.~Zhou}
\author{and the FRS-ESR, ILIMA, SPARC, and TBWD Collaborations}
\cortext[cor1]{Corresponding author}
\address[rvt]{Department of Physics, University of Istanbul, 34134 Istanbul, Turkey}
\address[cern]{CERN, 1211 Geneva 23, Switzerland}
\address[focal]{GSI Helmholtzzentrum f{\"u}r Schwerionenforschung, 64291 Darmstadt, Germany}
\address[jlu2]{II. Physikalisches Institut, Justus-Liebig Universit{\"a}t, 35392 Gie{\ss}en, Germany}
\address[saclay]{CEA, IRFU, SACM, Centre de Saclay, 91191 Gif-sur-Yvette, France}
\address[jlu1]{I. Physikalisches Institut, Justus-Liebig Universit{\"a}t, 35392 Gie{\ss}en, Germany}
\address[austria]{Stefan Meyer Institut f{\"u}r subatomare Physik, Austrian Academy of Sciences, 1090 Vienna, Austria}
\address[impcas]{Institute of Modern Physics, Chinese Academy of Sciences, Lanzhou 730000, P.~R.~China}
\address[fffff]{Institute of High Energy Physics, Chinese Academy of Sciences, Beijing 100049, P.~R.~China}
\address[triumf]{TRIUMF, Vancouver, British Columbia V6T 2A3, Canada}
\address[d]{Technische Universit{\"a}t M{\"u}nchen, 85748 Garching, Germany}
\address[e]{Helmholtz Institute Jena, 07743 Jena, Germany}
\address[guf]{J.~W.-Goethe-Universit{\"a}t, 60438 Frankfurt, Germany}
\address[ccrf]{Division of Radioisotope Research, CCRF, Niigata University, Niigata 950-8510, Japan}
\address[edinb]{School of Physics \& Astronomy, The University of Edinburgh, Edinburgh EH93FD, U. K.}
\address[saitama]{Department of Physics, Saitama University, Saitama University, Saitama 338-8570, Japan}
\address[tsukuba]{Institute of Physics, University of Tsukuba, Ibaraki 305-8571, Japan}
\address[niigata]{Department of Physics, Niigata University, Niigata 950-2181, Japan}
\address[kracow]{AGH University of Science and Technology, 30-059 Krakow, Poland}
\address[f]{Friedrich-Schiller-Universit{\"a}t Jena, 07743 Jena, Germany}
\address[beihang]{School of Physics \& Nucl. Energy Engineering, Beihang Univ., 100191 Beijing, P.~R.~China}
\address[riken]{RIKEN Nishina Center for Accelerator-Based Science, Wako 351-0198, Japan}
\address[spbu]{St. Petersburg State University, 198504 St. Petersburg, Russia}
\address[paris]{INSP, CNRS, Sorbonne Universit{\'e}, UMR 7588, 75005 Paris, France}
\address[surrey]{Department of Physics, University of Surrey, Guildford GU2 7XH, U. K.}
\address[emmi]{ExtreMe Matter Institute EMMI, 64291 Darmstadt, Germany}

\begin{abstract}
An experiment addressing electron capture (EC) decay of hydrogen-like $^{142}$Pm$^{60+}$ 
ions has been conducted at the experimental storage ring (ESR) at GSI. 
\red{The decay appears to be purely exponential and no modulations were observed.}
Decay times for about 9000 individual EC decays have been measured by applying the single-ion decay spectroscopy method.
Both visually and automatically analysed data can be described by a single exponential decay 
with decay constants of 0.0126(7) s$^{-1}$ for automatic analysis and 0.0141(7) s$^{-1}$ for manual analysis. 
If a modulation superimposed on the exponential decay curve is assumed, 
the best fit gives a modulation amplitude of merely 0.019(15),
which is compatible with zero and by 4.9 standard deviations smaller than in the original observation which had an amplitude of 0.23(4).
\end{abstract}


\end{frontmatter}



\section{Introduction}
\label{}
Highly charged ions (HCI) offer unrivalled opportunities for precision weak decay studies \cite{Litvinov2011,Bosch2013,Atanasov2015}.
In contrast to neutral atoms with complicated effects of many bound electrons \cite{Bambynek}, 
nuclei with none or just a few orbital electrons represent ``clean'' quantum mechanical systems.
The decay properties of HCIs can significantly be different from the ones known in neutral atoms 
\cite{Irnich1995,Folan1995,Phillips1989,Litvinov2003,Liu2006,Sun2007,Litvinov2007,Patyk2008,Ivanov2008a,Winckler2009a,Chen2010,Siegen2011,Atanasov2012}.
A straightforward example is the orbital electron capture decay which is simply disabled in fully ionised atoms.
Furthermore, HCIs enable investigations of exotic weak decay modes that are strongly suppressed or even forbidden in neutral atoms \cite{Litvinov2011,Bosch2013}.
A striking example of such a decay mode is the bound-state beta decay 
\cite{Daudel1947,Bahcall1962,Jung1992,Bosch1996,Ohtsubo2005,Kurcewicz2010,Pavicevic2018}.

An essential prerequisite for weak decay studies of radioactive HCIs is their production 
in a defined high atomic charge state and their controlled storage in this charge state over a sufficiently long period of time.
The facilities at the GSI Helmholtz Center for Heavy Ion Research in Darmstadt are ideally suited for weak decay studies of HCIs.
The GSI accelerator complex consists of three key elements \cite{Franzke2008}: the heavy-ion synchrotron SIS18 \cite{Blasche1985}, 
the projectile fragment separator FRS \cite{Geissel1992} and the heavy-ion cooler storage ring ESR \cite{Franzke1987}.
Except for a few exceptions, all experiments on radioactive decays of HCIs were conducted at \red{the} ESR, 
see Refs. \cite{Atanasov2017,Faestermann2015,Singh2018} and references cited therein.

An intriguing observation was published in 2008 
where modulated electron capture (EC) decays of 
hydrogen-like $^{140}$Pr$^{58+}$ and $^{142}$Pm$^{60+}$ ions were measured in \red{the} ESR \cite{Litvinov2008}.
Both, $^{140}_{~59}$Pr and $^{142}_{~61}$Pm nuclei can decay via the three-body $\beta^+$ 
and two-body EC pure Gamow-Teller $(1^+\rightarrow0^+)$ transitions to stable $^{140}_{~58}$Ce and $^{142}_{~60}$Nd nuclei, respectively \cite{NNDC}.

The modulated decay constant can be approximated by
\begin{equation}
\tilde\lambda_{EC}=\lambda_{EC}\cdot[1+a\cdot cos(\omega{t} + \phi)], 
\label{equation1}
\end{equation}
with the unmodified EC decay constant $\lambda_{EC}$ and an amplitude $a$, an angular frequency $\omega$ and a phase $\phi$ of the modulation.
Very similar frequencies $\omega = 0.89(1)$ rad s$^{-1}$ and 0.89(3) rad s$^{-1}$ 
as well as amplitudes $a=0.18(3)$ and 0.23(4) and quite different phases 0.4(4) rad and -1.6(5) rad
were measured for $^{140}$Pr$^{58+}$ and $^{142}$Pm$^{60+}$ ions, respectively \cite{Litvinov2008}.
The averaged amplitude is $\langle a \rangle=0.20(2)$.
 
The peculiarity of that experiment was that only a very few ions (on average 2 ions) were stored simultaneously in each injection of the ions into \red{the} ESR.
For more details see section \ref{experiment}.
This is the so-called {\it single-ion decay spectroscopy} method.
The advantage of this approach is that each individual EC decay was identified and its time was accurately determined.
The clear disadvantage was the very limited accumulated counting statistics.
In the first experiment merely 2650 (2740) EC decays were measured from 7102 (7011) injections into \red{the} ESR 
of $^{140}$Pr$^{58+}$ ($^{142}$Pm$^{60+}$) ions, respectively.
Because of the small counting statistics the statistical significance of the observed effect was not very high (about 3$\sigma$). 

The observation of the modulated weak decay caused an intensive controversial discussion in the literature.
For a non-exhaustive list the reader is referred to 
Refs.~\cite{Lipkin2009,
Lipkin2008,
Giunti2008,
Burjhardt2008,
Kienert2008a,
Ivanov2008,
Ivanov2008b,
Xing2008,
Faber2008,
Vetter2008,
Faestermann2009,
Lambiase2008,
Winckler2009b,
Kienle2009,
Merle2009,
Cohen2009,
Ivanov2009,
Ivanov2009b,
Kleinert2009,
Giunti2009,
Kienle2010x,
Lipkin2010,
Lipkin2010b,
Flambaum2010,
Ivanov2010a,
Ivanov2010b,
Pavlichenkov2010,
Gal2010,
Wu2010,
Merle2010,
Giunti2010,
Kienle2011,
Wu2011,
Krainov2012,
Giacosa2013,
Lambiase2013,
Peshkin2014,
Peshkin2015,
Alavi2015,
Gal2016}
and references cited therein.
It was therefore essential to experimentally confirm the observation on a higher statistical level.
Furthermore, it \red{was} important to identify physical quantities responsible for the modulation parameters.

Several attempts were performed by selecting different mass numbers, 
different charge states and different decay modes for ions in different experiments \red{at the ESR} since 2008.
However, except for the case of hydrogen-like $^{122}$I$^{52+}$ ions \cite{Faestermann2015,Kienle2009,Kienle2009b}, 
no statistically significant modulated decays were observed.
Although in the case of $^{122}$I$^{52+}$ ions an indication of a considerable modulation has been observed \cite{Winckler2008c,Kienle2011,Kienle2009b,Kienle2010,Kienle2010x},
the signal-to-noise characteristics of the obtained spectra had questioned the overall quality of the measured data.
Different data analyses did not converge and the final experimental results remained unpublished.
Therefore, it has been decided to repeat the very first experiment on one of the originally used hydrogen-like ion species.
The choice was made to use $^{142}$Pm$^{60+}$ ions.
In order to increase the reliability of the measured data, a significant effort has been put into the improvement 
of the detectors and the data acquisition system (see section \ref{experiment}).

The experiment was repeated in 2010~\cite{Kienle2013}.
A newly developed detector system (see section \ref{experiment}) has been employed together with the older system used in Ref. \cite{Litvinov2008}.
Altogether 17460 injections into \red{the} ESR of on average four parent $^{142}$Pm$^{60+}$ ions were done.
In total 8665 EC decays were recorded.
No significant modulation was observed in this entire data set~\cite{Kienle2013}.
However, a technical issue has been identified which might have caused a considerable systematic uncertainty.
In order to determine the decay time, it is essential to know the production time of each ion.
This requires that the ring is emptied before the fresh ions are injected.
Several indications of remaining ions from the last measurement cycles were documented during the experiment,
which indicated that there was a systematic problem with the employed emptying procedure~\cite{Piotrowski2014}. 
Under such conditions the determination of the decay times relative to the time of ion production would become impossible.
This could strongly influence the measurement results. 
Therefore, a differential analysis of the data has been done.
A long series of 3594 EC decays in 7125 consecutive injections was established.
A fit using Eq.~(\ref{equation1}) of these 3594 EC decays 
indicated the presence of a modulation with amplitude $a=0.107(24)$, angular frequency $\omega=0.884(14)$ rad s$^{-1}$, and phase $\phi=2.35(48)$ rad~\cite{Kienle2013}.
Striking was the angular frequency which was in excellent agreement with the one measured in the original experiment (see Table ~(\ref{jlab1})).
Whereas the different amplitudes might be due to the technical reason mentioned above,
the different phases could not been explained.

Inconsistency of results from performed experiments questioned the validity of the original observation.
A dedicated EMMI (ExtreMe Matter Institute) Rapid Reaction Task Force was called together in July 2014 \cite{EMMI2014} 
to thoroughly discuss all aspects of all performed experiments as well as published and unpublished data.
As a result, a recommendation was made to GSI management board to repeat the experiment 
under conditions as close as possible to the ones during the \red{very} first experiment \red{reported in 2008 \cite{Litvinov2008}}.

The new experiment, addressing EC decay of hydrogen-like $^{142}$Pm$^{60+}$ ions was conducted in Autumn 2014.
The state of the art detector equipment has been used offering significantly increased sensitivity as compared to the measurement in Ref. \cite{Litvinov2008}.
In this work we report the results of this measurement at \red{the} ESR.

\section{Experimental Method}
\label{experiment}
The experiment involved all major accelerator structures of GSI, namely 
universal linear accelerator (UNILAC), heavy ion synchrotron (SIS18), fragment separator (FRS) and experimental storage ring (ESR).
In order to avoid any possible distortions or influences, there were no other experiments running in parallel at \red{the} SIS18.

As in the previous experiments, the primary beam of $^{152}$Sm has been used.
Several pulses of primary beams were accumulated and electron cooled in \red{the} SIS18.
The beam was then accelerated to relativistic energy of $E_{\rm SIS18}=607.4$~MeV/u 
and extracted towards the production target placed at the entrance of \red{the} FRS.
In the employed fast extraction scheme, the entire beam was guided out of \red{the} SIS18 within one revolution, that is within 1~$\mu$s.

Before reaching the target, the beam passed through a negligibly thin carbon window and a SEETRAM (SEcondary Electron TRAnsmission Monitor) detector 
which consists of one titanium foil of 10~$\mu$m thickness sandwiched between two aluminium foils of 14~$\mu$m thickness each~\cite{SEETRAM1992}. 
A 2511~mg/cm$^2$ thick $^9$Be was used as a target.
Different ion species were produced in projectile fragmentation reactions.
The fragments were kinematically focused in forward direction and entered \red{the} FRS.
Among them were the $^{142}$Pm ions of interest in different atomic charge states.
According to LISE++ \cite{lise1,lise2,lise3} and MOCADI \cite{MOCADI} calculations $^{142}$Pm ions emerged from the target with energies of about 458~MeV/u.
The target thickness was large enough to safely assume the equilibrium charge state distribution of the fragments.
According to calculations with the GLOBAL code \cite{Scheidenberger1998}, about 84\% of Pm ions exited the target as fully-stripped, bare nuclei.
Therefore, the FRS magnets until the middle focal plane were set such that the fully-ionised $^{142}$Pm$^{61+}$ ions are centred in the ion-optical system.
The extraction time of 1~$\mu$s represents the uncertainty of the creation time of $^{142}$Pm ions.

The daughter ions of the EC decay of the hydrogen-like $^{142}$Pm$^{60+}$ ions are bare $^{142}$Nd$^{60+}$ nuclei.
It was important to remove all contaminants that can produce $^{142}$Nd$^{60+}$ ions through various other channels \red{during the storage in the ESR}.
For instance, $^{142}$Nd$^{60+}$ can be produced from $^{142}$Nd$^{59+}$ via stripping the bound electron in the rest gas of \red{the} ESR.
Also, the presence of parent ions in other charge states should be excluded.
An energy degrader composed of a 737~mg/cm$^2$ aluminium disk and 256~$\mu$m niobium foil has been used at the middle focal plane of \red{the} FRS
to enable the $B\rho-\Delta E-B\rho$ separation method \cite{BrDEBr}, where $B\rho$ and $\Delta E$ stand for magnetic rigidity and atomic energy loss, respectively.
By selecting the fully-ionised $^{142}$Pm$^{61+}$ ions in the first half of \red{the} FRS, no $^{142}$Nd ions were transmitted to the degrader.
The usage of a niobium foil shall optimise the production of the hydrogen-like ions.
According to the GLOBAL code \cite{Scheidenberger1998}, about 10\% of Pm ions were in the hydrogen-like charge state after the Nb foil.
The second half of \red{the} FRS was set such that $^{142}$Pm$^{60+}$ ions are centred in the ion-optical system.
Nearly pure beams of $^{142}$Pm$^{60+}$ ions have been transmitted to and injected into \red{the} ESR.
The energy of the primary $^{152}$Sm beam has been selected such that $^{142}$Pm$^{60+}$ ions reached \red{the} ESR with an energy of $E_{\rm ESR}=400$~MeV/u.
The calibration of all material thicknesses and the optimisation of beam injection into \red{the} ESR has been done with the primary beam.

\begin{figure*}[t!]
\centering
\includegraphics[angle=90,width=0.5\textwidth,angle=270]{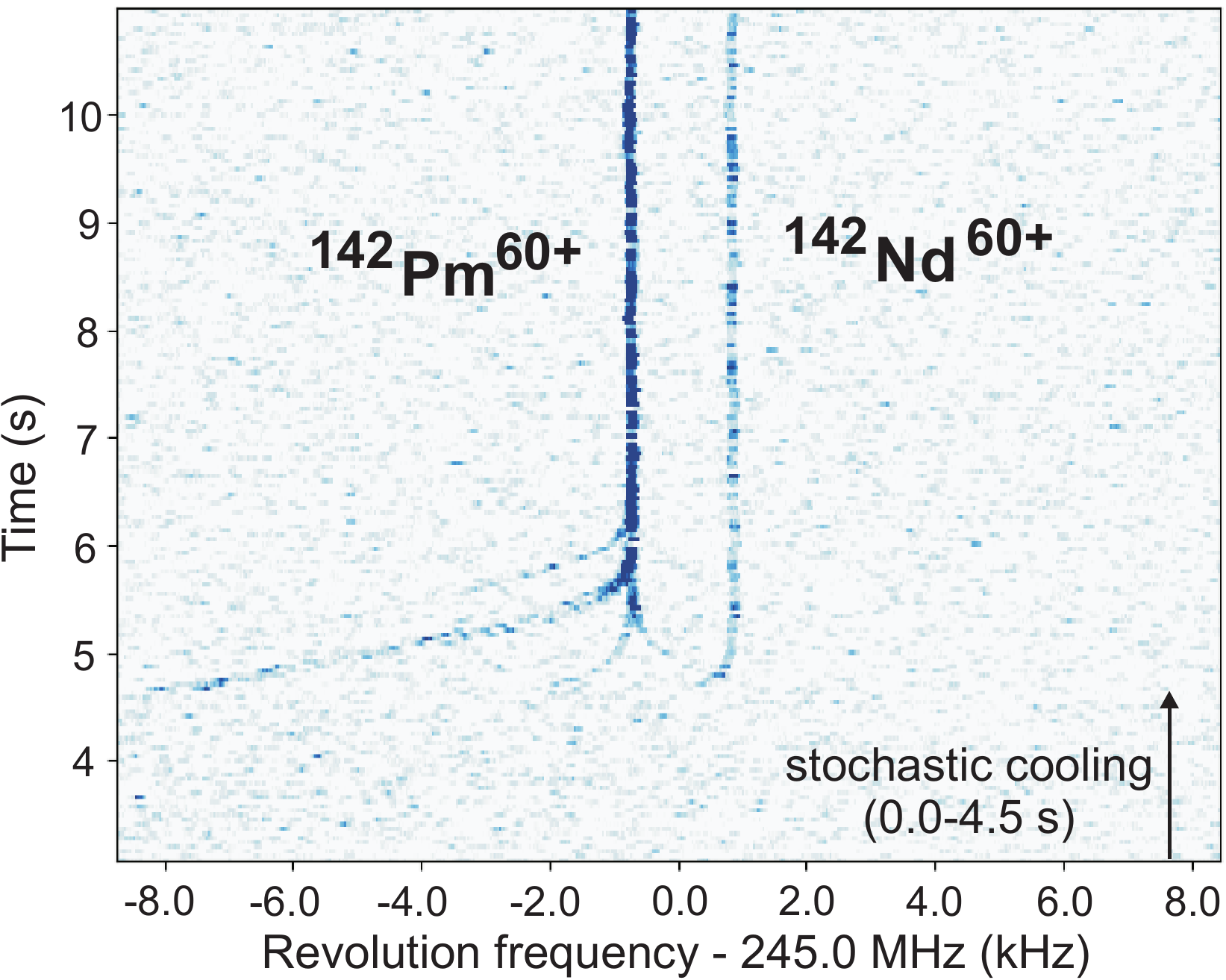}
\caption{Example of the measured traces of stored $^{142}$Pm$^{60+}$ and $^{142}$Nd$^{60+}$ ions in \red{the} ESR. 
More than six $^{142}$Pm$^{60+}$ were stored in this example.
Also a single EC-decay daughter ion, $^{142}$Nd$^{60+}$, in a frequency-time after injection representation is present from the beginning.
The vertical scale is zoomed on the first 10 seconds of the measurement to illustrate the duration of stochastic cooling.
The injection of ions into \red{the} ESR occurs at 0 seconds.
The stochastic cooling is operated from 0 to about 4.5 seconds.
The electron cooling is operated all the time at unchanged parameters without interruptions.
Cooling individual ions by the cooler electrons is clearly seen from 4.5 seconds up to about 6 seconds.
\label{stochastic}}
\end{figure*}

The $^{142}$Pm$^{60+}$ ions were cooled in \red{the} ESR by employing stochastic \cite{Nolden2004} and electron \cite{Steck2004} cooling.
The former method operates at a fixed ion velocity corresponding to $E_{\rm ESR}=400$~MeV/u.
In this experiment the stochastic cooling could be optimised such that its operation time was about 4.5 seconds, see Figure~\ref{stochastic}.
The electron cooling was continuously switched on with unchanged parameters.
In this experiment the electron current of 250~mA and the acceleration potential of 219850~V have been used.
These parameters were optimised for the cooling electrons to match the velocity of the ions after the stochastic cooling.
The velocity spread of the cooled ions was about $\Delta v/v\approx 5\cdot10^{-7}$.
The ions coasted in the ring with a velocity $\beta=v/c=0.71$ corresponding to relativistic Lorentz factor of $\gamma=1.43$.
The acceptance of \red{the} ESR has been reduced by inserting copper scrapers into its aperture to remove any products of 
atomic charge exchange reactions or three-body beta decays of the ions of interest.

\begin{figure*}[t!]
\centering
\includegraphics[width=0.6\textwidth]{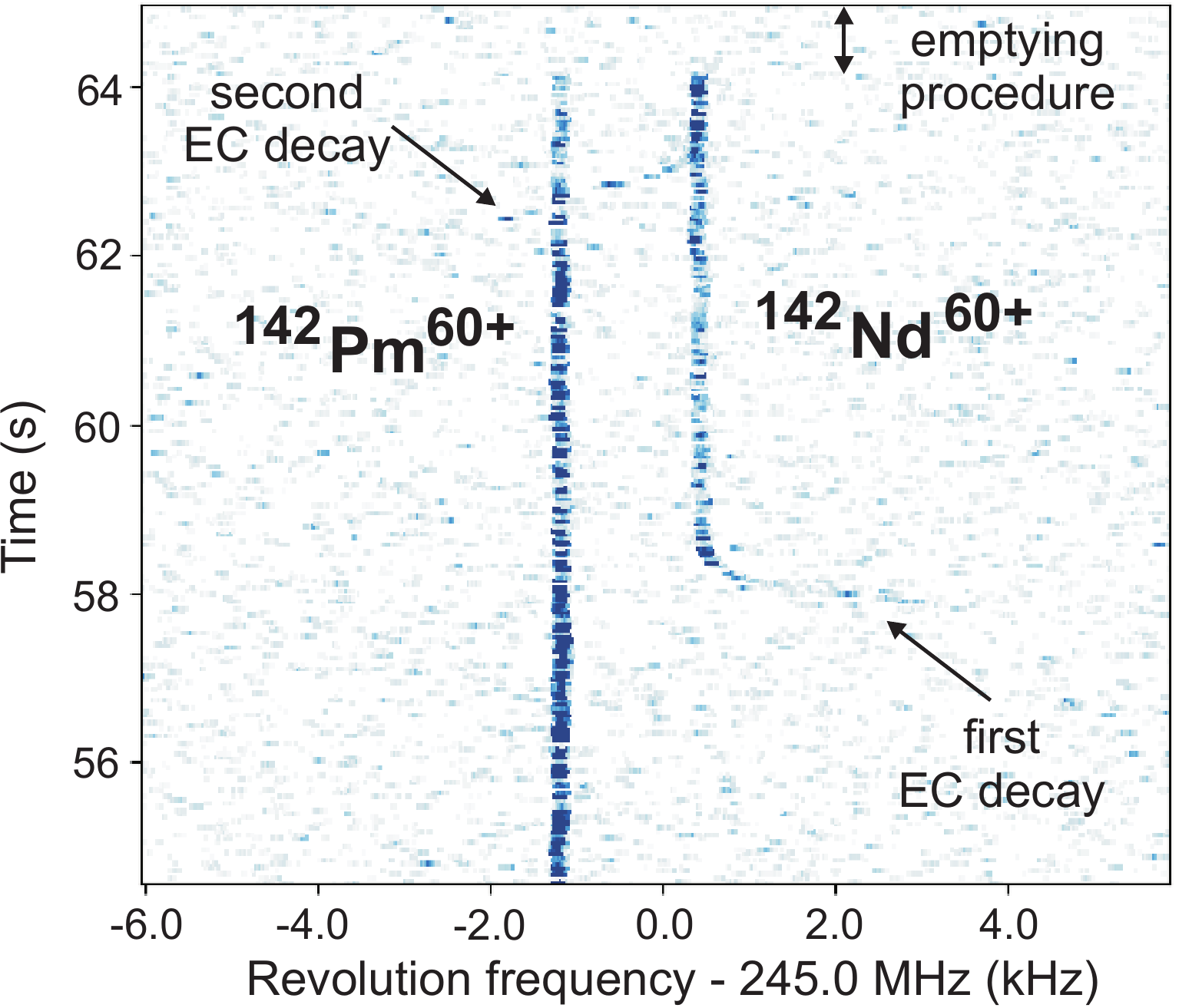}
\caption{Example of the measured EC decays of stored and cooled $^{142}$Pm$^{60+}$ ions. 
The vertical axis is zoomed from the measured range of $0-70$ seconds.
Two EC decays are clearly seen at about 58 and 63 seconds after the injection of the ions into \red{the} ESR.
As has been shown in \cite{Kienle2013}, neutrinos are emitted isotropically indicating that the stored $^{142}$Pm$^{60+}$ ions are unpolarised.
The longitudinal component of the recoil (due to the emitted neutrino) of the daughter ion is reflected by the frequency difference between
the frequency at which the daughter ion appears after the decay and its frequency when cooled by the electrons.
The tail at 58 seconds shows that the recoiling ion was slowed down by the electrons (to smaller revolution frequencies), 
which means that the neutrino was emitted in the direction opposite to the ion motion.
Vice versa is the case of the tail at 63 seconds.
The disappearance of both ion species at 64 seconds is due to the implementation 
of the kicker magnet pulse for safe and controlled emptying \red{the} ESR prior to the injection of newly produced $^{142}$Pm$^{60+}$ ions.
\label{trace}}
\end{figure*}

The observation time was set to 64 seconds, 
which is comparable to the expected halflife of $^{142}$Pm$^{60+}$ in the laboratory frame of about 56 seconds \cite{Winckler2009a}.
Afterwards an extraction kicker has been used to empty \red{the} ESR.
Additional 6 seconds were recorded after the kicker event to confirm the emptying of \red{the} ESR.
One example of the measured spectra is shown in Figure~\ref{trace}.
In order to ensure that the kicker modules function properly, their voltage versus time responses at each operation were stored on disk.
\redx{It has to be emphasised that the overall stability and reliability of the system has been significantly improved}. 
Only very few events have been observed when the kicker equipment failed \cite{Piotrowski2014}.

The decays of $^{142}$Pm$^{60+}$ ions were measured with {\it time-resolved Schottky mass spectrometry} \cite{Litvinov2004, Geissel2004, Litvinov2005, Geissel2007}.
Since the velocities of the ions were defined by the electrons of the cooler, their revolution frequencies reflected directly their mass-to-charge ratios.
A non-destructive Schottky detector was used to continuously monitor the frequencies of stored ions.
In the first experiment \red{only} a capacitive parallel-plates detector has been used \cite{Schaaf}.
\red{When using this capacitive detector, \redx{an averaging over 384~ms  
was required to detect single stored HCIs, see \cite{Litvinov2008} for more details.}
Although the detection of the first EC decay was sufficiently accurate, 
the low signal-to-noise ratio could lead to systematic errors in the identification of the second and further decays occurring within the same storage period.
This effect is amplified by the fact, that Schottky signal fluctuations increase with increasing the signal \redx{amplitude} \cite{Schaaf}.
Therefore, significant efforts were \redx{made} to increase the sensitivity of the detector.
A dedicated cavity-based resonant Schottky detector was developed for our experiments \cite{Nolden2011}.
This resonant detector has signal-to-noise characteristics which are at least one order of magnitude higher than the capacitive detector.
The summary of experimental parameters is given in Table~\ref{tabsum}.}

The principle of the cavity-based Schottky detector is similar to that of a transformer.
The revolution frequencies of the ions in \red{the} ESR were about 1.96 MHz.
The detector has its maximal sensitivity at the $125^{th}$ harmonic of the revolution frequency at around 245 MHz.
The signal acquired by the detector was amplified by low-noise pre-amplifier 
(BNZ1035, gain 39 dB), amplifier (ZKL1R5+, gain 40 dB), passed through low- and high-pass filters, and then
transported from \red{the} ESR to the main control room located about 380~m away.
The details of the signal chain and specifications of the employed high frequency parts can be found in \cite{Shahab}.

In the control room the signal was split and put into several data acquisition systems.
The main recording system in this experiment was composed of two real-time spectrum analysers from Tektronix, RSA 5103A and RSA 5126A, 
which were set to monitor 10~kHz and 15~kHz frequency bandwidths ($125^{th}$ harmonic) around the central frequency of the cooled $^{142}$Pm$^{60+}$ ions, respectively.
The online monitoring was done with the real-time spectrum analyser Tektronix RSA 3303B,
which was used instead of the older \red{Sony-Tektronix} employed as the main acquisition system in \cite{Litvinov2008}.
Another signal was taken to the New Time CAPture (NTCAP) system which is a broad-band real-time recording system developed by the collaboration \cite{Trageser2015,Trageser2018}.
The RSA devices can record a very limited frequency bandwidth of a few kHz.
In contrast, the NTCAP system is capable to monitor several MHz bandwidth \cite{Trageser2015}.
\red{Different acquisition devices employed in the three experiments are summarised in Table~\ref{tabsum}.}

The parameters of the spectra recorded in \red{the very first experiment by the Sony-Tektronix device} 
(frequency resolution, windowing function of the Fourier transform) 
had to be fixed prior to the measurements. 
Different to this, the data acquired in 2010 and 2014 experiments are in the raw format 
allowing for flexible optimisation of parameters in the production of spectra.
Three-dimensional plots in this work are made from data acquired with RSA 5103A device.
Each file contains \red{1.7 million \redx{complex sample} points}
acquired with 24.4 kilosamples per second.
The spectra are produced with multi-taper digital transform without windowing. 
The full measurement cycle of 70.0 seconds is represented by 1669 frequency spectra.
The frequency and time resolutions are 23.84 Hz/channel and 41.94 ms/channel, respectively. 

Figure \ref{trace} shows an example of the measured EC decay.
The mass of the ion changes in the decay and (if the number of particles is not large) the decay event is unambiguously observed by the 
reduction of the Schottky signal 
at the frequency corresponding to the parent ion and correlated in time with the
increase of the signal at the daughter-ion frequency. 
Due to the recoil momentum from neutrino emission, the longitudinal velocity is a bit off the cooling velocity, leading to a ``cooling tail''.

\section{Results}

 
The data analysis has been performed by several independent groups.
Each group has inspected each recorded 70-s file visually.
The following information has been collected from each file: 
\begin{itemize}
\item{Number of injected ions. This has been done by zooming onto the first few seconds, as illustrated in Figure~\ref{stochastic}, and counting the number of ``cooling tails'';}
\item{The decay time of each EC-decay. This has been done by zooming onto each decay event (see Figure~\ref{trace}). The time bin was taken at which the onset of the Schottky signal of the daughter ion was observed. The accuracy of such time determination is a few time bins corresponding to a few ten milliseconds;}
\item{The length of the corresponding ``cooling tail'' in Hertz for each EC-decay.}
\end{itemize}
All three individual visual analyses of the 2014 data set provided consistent results.

Furthermore, a dedicated automatic analysis program has been applied.
The algorithm is described in detail in \cite{Buehler2016}.

\begin{figure*}[hbtp]
\centering
\includegraphics[width=0.5\textwidth, angle=90]{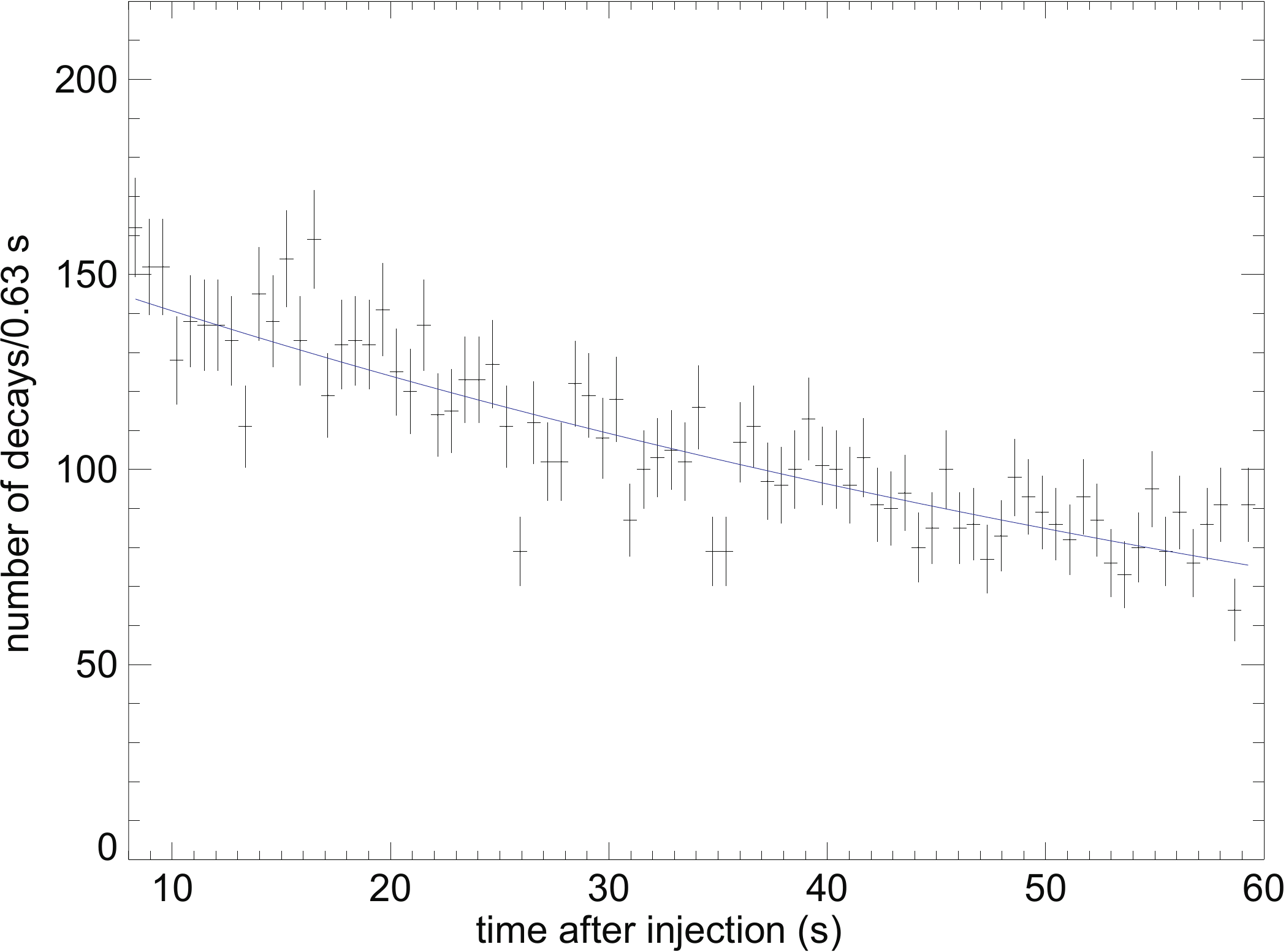}
\includegraphics[width=0.5\textwidth, angle=90]{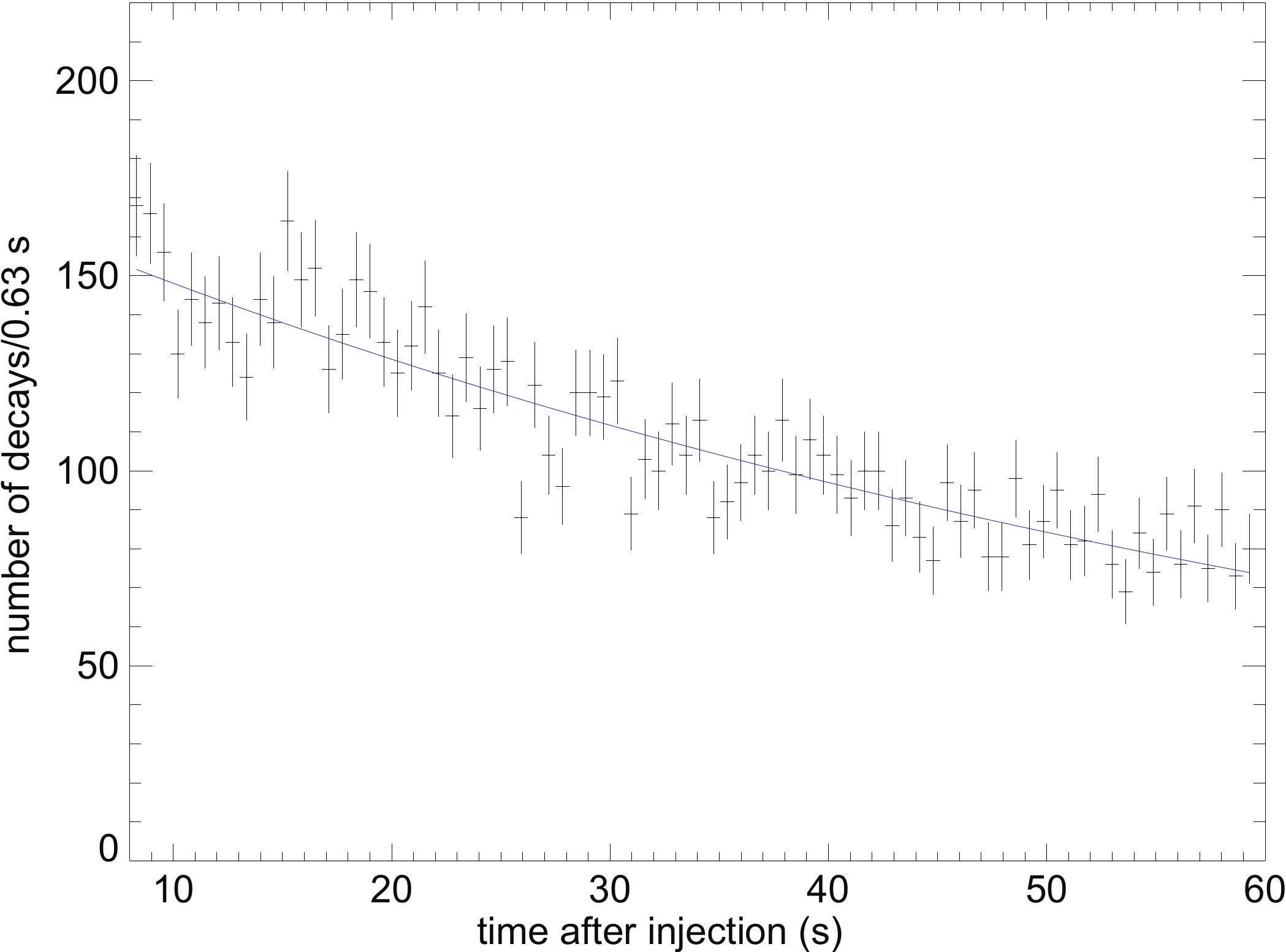}
\caption{(Top) Number of EC-decays per 0.63 s as determined in the automatic analysis (8839 EC decays in total). The data points are fitted with a pure exponential function (solid line). 
(Bottom) Same as (Top) but for one of the manual analyses (9001 EC decays in total), see text. 
\label{results1}}
\end{figure*}

About 9000 EC decays of parent $\ce{^{142}Pm^{60+}}$ ions have been analysed. 
The histograms obtained in the automatic (8839 EC decays) and in one of the manual analyses (9001 EC decays) are shown in Figure~\ref{results1}.
\begin{table*}
\caption{\label{tabsum}
\red{
Summary of key characteristics of the three experiments.}}
\footnotesize
\centering
\begin{tabular}{@{}lcccccc}
Year& Ion & Number of decays & Detector & Data Acquisition & Analysis & Ref.\\
\hline
2008&$^{140}$Pr$^{58+}$& 2650 & capacitive & Sony-Tektronix & manual & \cite{Litvinov2008}\\
2008&$^{142}$Pm$^{60+}$& 2740 & capacitive & Sony-Tektronix & manual & \cite{Litvinov2008}\\
2010&$^{142}$Pm$^{60+}$& 8665 & capacitive, resonant & Sony-Tektronix, RSA3303B  & manual, automatic& \cite{Kienle2013}\\
2010$^*$&$^{142}$Pm$^{60+}$& 3594 & capacitive, resonant & Sony-Tektronix, RSA3303B  & manual, automatic& \cite{Kienle2013}\\
2014&$^{142}$Pm$^{60+}$& 9001 & resonant & RSA5103A, RSA5126A, NTCAP  & manual, automatic& this work\\
\hline
\end{tabular}\\
\flushleft{$^*$ subset of data, see text\\}
\end{table*}
\normalsize

\begin{table*}
\caption{Results of the $\chi^2$ analysis of the data. 
\label{jlab1}$\lambda$ values and oscillation parameters obtained from different experiments on two-body electron capture decay of 
hydrogen-like $^{142}$Pm$^{60+}$ ions according to their years.
The values are in the laboratory system (Lorentz factor $\gamma=1.43$).
Labels ``e'' and ``m'' in the second column indicate the results for pure exponential or exponential plus a modulation fits, respectively.}
\footnotesize
\centering
\begin{tabular}{@{}lcccccc}
Year& &{$\lambda$}($\Delta${$\lambda$}) [s$^{-1}$]&a($\Delta${a})&{$\omega$}($\Delta${$\omega$}) [rad s$^{-1}$]&{$\phi$}($\Delta${$\phi$}) [rad]&Ref.\\
\hline
2008&e&0.0170(9)&-&-&-&\cite{Litvinov2008}\\
2008$^*$&m&0.0224(42)&0.23(4)&0.885(31)&-1.6(5)& \cite{Litvinov2008}\\
2008$^{**}$&e&0.0124(2)&-&-&-&\cite{Winckler2009a}\\

2010&m&0.0130(8)&0.107(24)&0.884(14)&+2.35(5)&\cite{Kienle2013}\\
2014$^a$&e&0.0126(7)&-&-&-&this work\\
2014$^m$&e&0.0141(7)&-&-&-&this work\\
2014$^m$&m&0.0141(9)&0.019(15)&1.04(26)&-3.1(2)&this work\\
\hline
\end{tabular}\\
\flushleft{$^*$ the results of the fit for the data until 33 seconds after injection\\
$^{**}$ the results for measurements with several thousands stored ions\\
$^a$ the results of the automatic analysis\\
$^m$ the results of the manual analysis}
\end{table*}
\normalsize
 
\begin{figure*}[hbtp]
\centering
\includegraphics[width=0.5\textwidth]{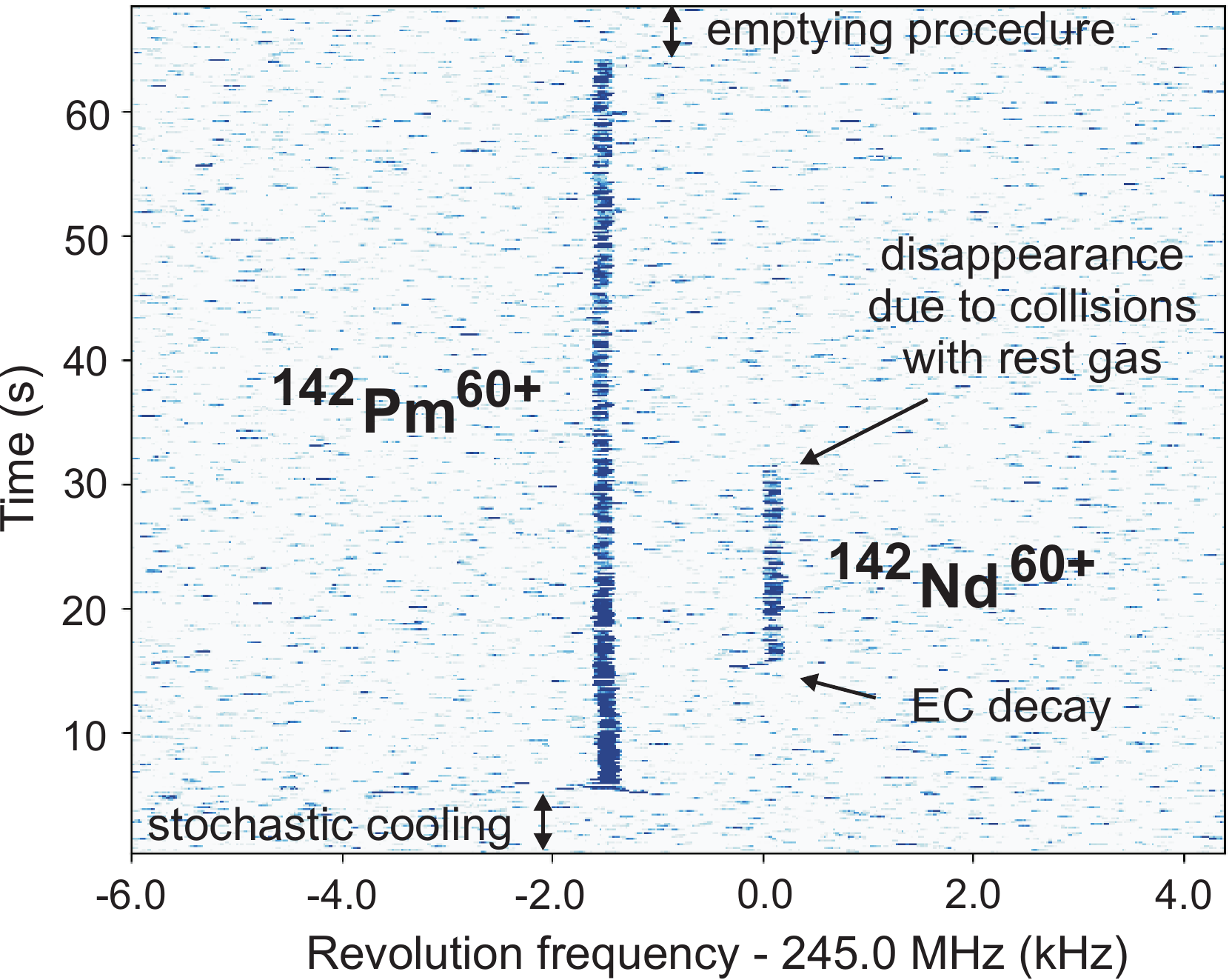}
\caption{Disappearance of a stable $\ce{^{142}Nd^{60+}}$ ion which is due to non-radioactive losses of the ions from the ring. The entire measurement cycle is shown.
\label{lambda_loss}}
\end{figure*}

Already a brief inspection of Figure~\ref{results1} indicates that no significant modulation of the number of decays in time is observed.
The data have been analysed according to two decay models.
The first one is the strictly exponential decay 
\begin{equation}
\frac{dN_d(t)}{dt}={\lambda_{EC}}{N_{M}(0)}e^{-\lambda{t}}\label{eq2},
\end{equation}
and the second one assuming a modulated $\tilde\lambda_{EC}$ as given by Eq.~(\ref{equation1}).
Here, 
$dN_{d}(t)$ is the number of daughter ions observed at time $dt$, 
$N_{M}(0)$ number of parent ions at time $t=0$ seconds,
$\lambda_{EC}$ the EC decay constant, 
$\lambda$ the total decay constant $\lambda$ = $\lambda_{\beta^+}$+$\lambda_{EC}$+$\lambda_{loss}$. 
The $\lambda_{\beta^+}$ is the decay constant of the three-body beta decay, $\lambda_{EC}$ the EC decay constant, and $\lambda_{loss}$ is the decay constant due to unavoidable (non-radioactive) losses of the ions from the ring.
The latter losses can be estimated through disappearance of stable daughter $\ce{^{142}Nd^{60+}}$ ions. 
Only a few such decays have been observed (see, e. g., Figure~\ref{lambda_loss}), which indicates that the $\lambda_{loss}$ constant can be neglected.

The results of the standard $\chi^2$ minimisation are presented in Table~\ref{jlab1} 
together with results from previous experiments taken from \cite{Litvinov2008, Kienle2013}.

{
In addition to the analysis based on the $\chi^2$ minimisation, 
an approach based on Bayesian statistics has been applied. 
Differently from criteria based on the comparison of $\chi^2$ (or likelihood) values to decide in favour or against one model, 
the Bayesian method allows for directly assigning a probability value to each model.
The model probability is based on the computation of the \textit{Bayesian evidence}.
The Bayesian evidence, also called \textit{marginal likelihood} or \textit{model likelihood} is calculated from the integral of the likelihood function over the parameter space.
For more details, the reader is referred to, e. g., \cite{Sivia,Trotta2008,Trassinelli2017b}.
Differently from maximum likelihood or minimum $\chi^2$ methods, 
this approach is particularly adapted when multiple local maxima of the likelihood function are present, as in our case.
Moreover, Bayesian evidence naturally encodes \red{Ockham's} razor principle, penalising models that are unnecessarily too complex for generating the observed data. 

The computation of the evidence relative to the two models has been done using two independent approaches.
In the first method, the likelihood function is built from unbinned data and its integration is obtained 
with the VEGAS Monte Carlo algorithm by using the BAT package \cite{Caldwell2009} in a self-made root-based code \cite{Winckler2015,Winckler2015a}.  
The second method is based on the \texttt{Nested\_fit} program \cite{Trassinelli2017b}, 
which is based on the \textit{nested sampling} algorithm for the transformation of the $n$-dimensional integral
(with \textit{n} number of parameters of the model) into a one-dimensional integral \cite{Skilling2006} 
and in a home-made Monte Carlo sampling.
Here, the likelihood function is computed from data binned with interval width comparable to the instrumental resolution and assuming a Poisson distribution for each channel. 
The results from the two independent analyses are very similar.
The results obtained from Nested\_fit approach are presented in Table~\ref{tab:martino}.

\begin{table*}
\caption{\label{tab:evidence}  
Probability of the model with oscillation ($P(M_2)$), parameter values corresponding to the maximum of the likelihood functions and their 95\% confidence intervals (CI) 
from the analysis with Nested\_fit program.
The considered range for $\omega$ is $[0,7]$ rad s$^{-1}$.}
\footnotesize
\centering
\begin{tabular}{@{}lccccc}
Year    &  $\lambda$    (CI\ 95\%)  [s$^{-1}$]   &  a (CI\ 95\%)     & $\omega$ (CI\ 95\%)  [rad s$^{-1}$]    &$\phi$ (CI\ 95\%) [rad]   & $P(M_2)$\\
\hline
2008    & $0.0207 (0.0155-0.0250)$ & $0.20 (1.3\times10^{-3}-0.019)$            & $0.91 (0.33-6.59)$     & $5.91(0.15-6.20)$    & $47.5-85.1\%$  \\
2010    & $0.0140 (0.0120-0.0156)$ & $9.2\times10^{-2} (2.2\times10^{-4}-7.2\times10^{-2})$    & $0.89 (0.17-6.86)$    & $3.84(0.18-6.14)$     & $58.1-65.8\%$  \\
2014    & $0.0149 (0.0136-0.0157)$ & $5.3\times10^{-2} (3.9\times10^{-4}-3.6\times10^{-2})$    & $4.71 (0.40-6.65)$    & $4.71 (0.22-6.04)$    & $52.3-67.1\%$  \\

\hline
\end{tabular} \label{tab:martino}
\end{table*}
\normalsize



\section{Discussion}
The fit procedure assuming the model of Eq.~(\ref{equation1}) always results in nonzero modulation parameters.
In our context, we search for statistically significant modulation parameters consistent with previous observations \cite{Litvinov2008, Kienle2013}.

By inspecting the data in Table~\ref{jlab1} it can be concluded that 
no significant modulation is observed at the previously reported modulation frequency of $\omega\approx0.89$ rad s$^{-1}$.
The largest modulation amplitude $a=0.019(15)$ corresponds to the modulation frequency of $\omega=1.04(26)$ rad s$^{-1}$
and deviates by 4.9$\sigma$/3.1$\sigma$ from the amplitudes reported in \cite{Litvinov2008} and \cite{Kienle2013},
respectively.

As well, the results of the Bayesian analysis indicate no significant modulation, see the $P(M_2)$ column in Table~\ref{tab:martino},
and the obtained best parameter values are very different from the previously reported ones.
\red{Furthermore, our new Bayesian analysis has been applied to the data acquired in previous 2008 \cite{Litvinov2008}  and 2010 \cite{Kienle2013} experiments. 
\redx{Presence of modulated EC decays could not be confirmed in past measurements.}
The probabilities assigned to the model with and without modulation (\red{Table}~\ref{tab:martino}) are found \redx{to be} similar (close to 50\%).}
The only strong indication of the presence of a modulation in the data is obtained 
considering the three data sets (from 2008, 2010 and 2014) simultaneously 
and considering three functions like Eq.~(\ref{equation1}) for each data set with free amplitudes and phases 
but with a common value of $\omega$ (Table~\ref{tab:martino_set}).
In this case, a probability of $P(M_2)=95.0-99.85\%$ is assigned to the model with modulation, 
corresponding to a p-value of $0.0034-5.5\times10^{-5}$ and $2.93-4.03$ $\sigma$ \cite{Gordon2007}, 
where the uncertainties are from the Bayesian evidence computation.
Because of the same experimental apparatus, different values of phase between 2010 and 2014 experiments are, however, unlikely. 
When the same value of $a$ and/or $\phi$ are imposed for these data sets, the models with and without modulation are equally probable.

\begin{table*}
\caption{\label{tab:evidence}  
Probability of the model with oscillation ($P(M_2)$), parameter values corresponding to the maximum of the likelihood functions and their 95\% confidence intervals (CI) 
from the analysis of all data sets at the same time with Nested\_fit program. 
The considered range for $\omega$ is $[0,7]$ rad s$^{-1}$.}
\footnotesize
\centering
\begin{tabular}{@{}lccc}
Constraints    & $\lambda$    (CI\ 95\%)     [s$^{-1}$] &{$\omega$} (CI\ 95\%)  [rad s$^{-1}$]    &  $P(M_2)$\\
\hline
None    &    $0.0144 (0.0142-0.0146)$ &  $0.911 (0.908-0.975)$     & $95.01-99.85\%$  \\
$\phi$ locked for data sets from 2010 and 2014	& $0.0143 (0.0137-0.0150)$ & $0.906 (0.507-6.65)$     & $44.48-85.59\%$  \\
$\phi$ and $a$ locked for data sets from 2010 and 2014& $0.0144 (0.0138-0.0153)$ &  $0.929 (0.363-6.71)$     & $41.40-62.82\%$  \\
\end{tabular}  \label{tab:martino_set}
\end{table*}
\normalsize


The question on the origin of the 20\% modulation observed in \cite{Litvinov2008} remains unanswered.
\red{In two datasets containing about 9000 EC decays, namely the full datasets from 2010 and from this work, the modulation amplitude is negligibly small.
Thus, the modulations were observed only in datasets with significantly smaller statistics.}

Several differences between the \red{very first experiment} reported in \cite{Litvinov2008} and this work can be mentioned:
\begin{itemize}
\item{The \red{Sony-Tektronix} data acquisition system employed in the first experiments could not be maintained due to its age. 
Furthermore, only the capacitive Schottky detector was available at that time. 
It is impossible to study whether the older system could cause, though unlikely, artefacts in the data.
The new data acquisition solutions \red{(see Table~\ref{tabsum})} as well as
the new \red{resonant Schottky} detector offer orders of magnitude increased performance of the overall system; 
}
\item{The number of injected ions in the latest (2014) experiment was larger than in the first experiments. Often more than 6-8 ions were stored.
\red{Therefore, some of the late EC-decays might have been missed in the manual analysis,
in which the decay is identified through the detection of a ``cooling tail'' (see Figure~\ref{stochastic})}. 
An example of such an EC-decay is illustrated in Figure~\ref{notail}. 
These decays correspond to the emission of the neutrino in transversal direction thus leaving the longitudinal velocity of the ion nearly unchanged. 
If such a decay occurs after several daughter ions are already produced, it might be missed in manual analyses, though not in the automatic analysis.
By comparison of different analyses one can conclude, that the missing decays have little influence on the general behaviour of the \red{decay curves} shown in Figure~\ref{results1}
except for the different decay constants resulting from pure exponential fits; 
}
\item{Some ``cooling tails'' could be due to a longitudinal momentum transfer in collisions of ions of interest with rest gas molecules. In such cases a tail on the low-frequency side can be observed. If the number of stored ions in a frequency peak is not high (below 3-5 ions) than the EC decay can unambiguously be identified by the correlated decrease of the intensity of the parent ions and an increase of intensity of the daughter ions. 
On the one hand, if the number of ions is larger, then some of such tails could erroneously be identified in manual analyses as being from the EC-decay.
On the other hand, the automatic analysis takes the changes of intensities into account and should discard such cases.}
\end{itemize}

\red{As compared to the experiment performed in 2010, the quality of the present data was somewhat lower.} 
For a single particle, the obtained signal-to-noise ratios \red{were} 26 and 9 for 2010 and 2014 experiments, respectively. 
\red{The reason for this reduced signal-to-noise ratio in 2014 might have been the larger number of acquisition devices, 
which required the \redx{additional} splitting of the signal from the detector.}
Furthermore, a large number of un-identified un-cooled particles were present in the storage ring. 
One such example is illustrated in Figure~\ref{travellers}. 

\begin{figure*}[hbtp]
\centering
\includegraphics[width=0.5\textwidth]{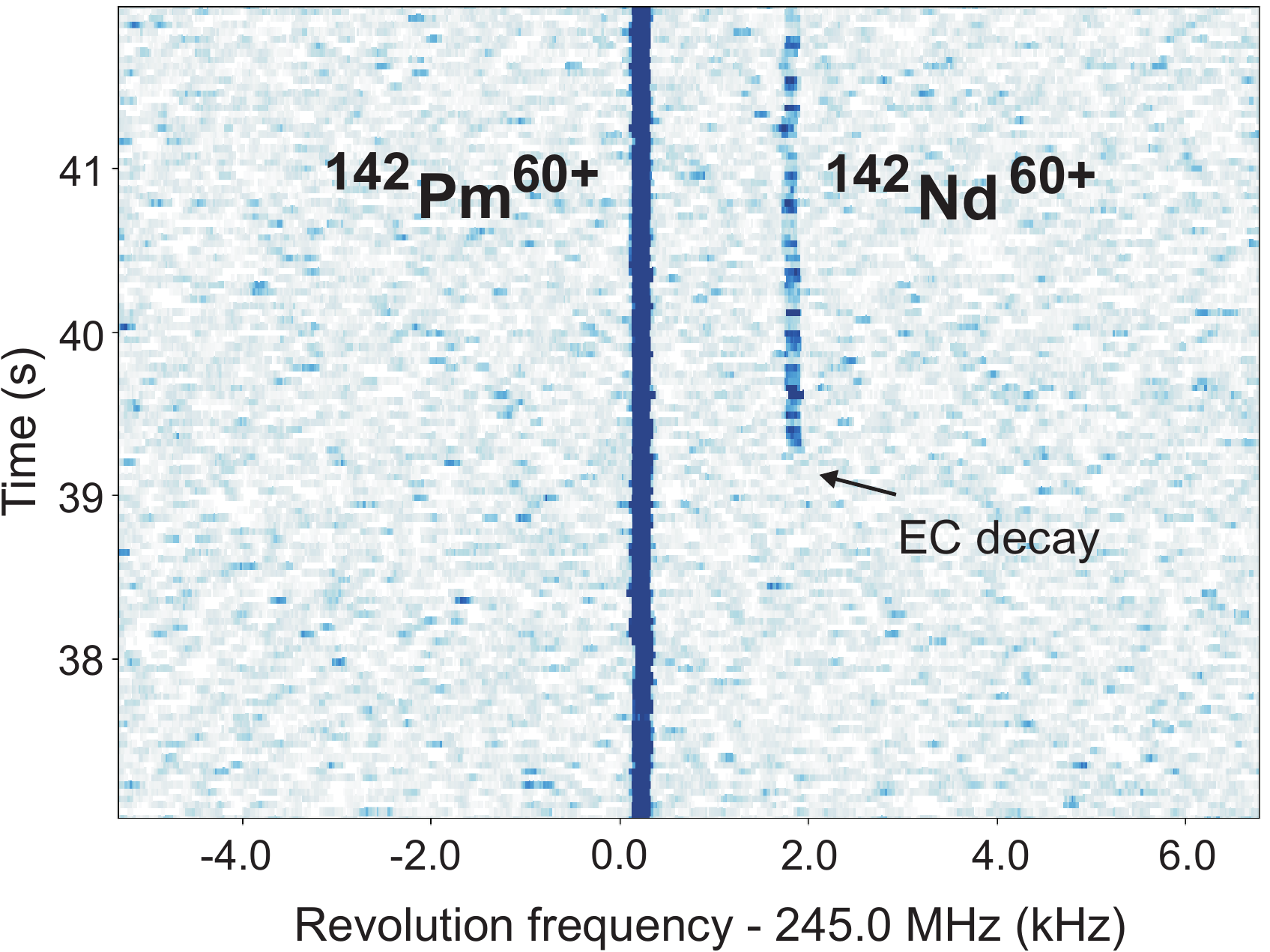}
\caption{Traces of $^{142}$Nd$^{60+}$ daughter ions without a significant tail. 
If several daughter ions are present at the time of such decay, the identification of the latter in a manual analysis is complicated.
The automatic analysis however finds decays independent of the occurrence of a cooling tail and thus should be less prone to overlooking such cases.
\label{notail}}
\end{figure*}


\begin{figure*}[hbtp]
\centering
\includegraphics[width=0.5\textwidth]{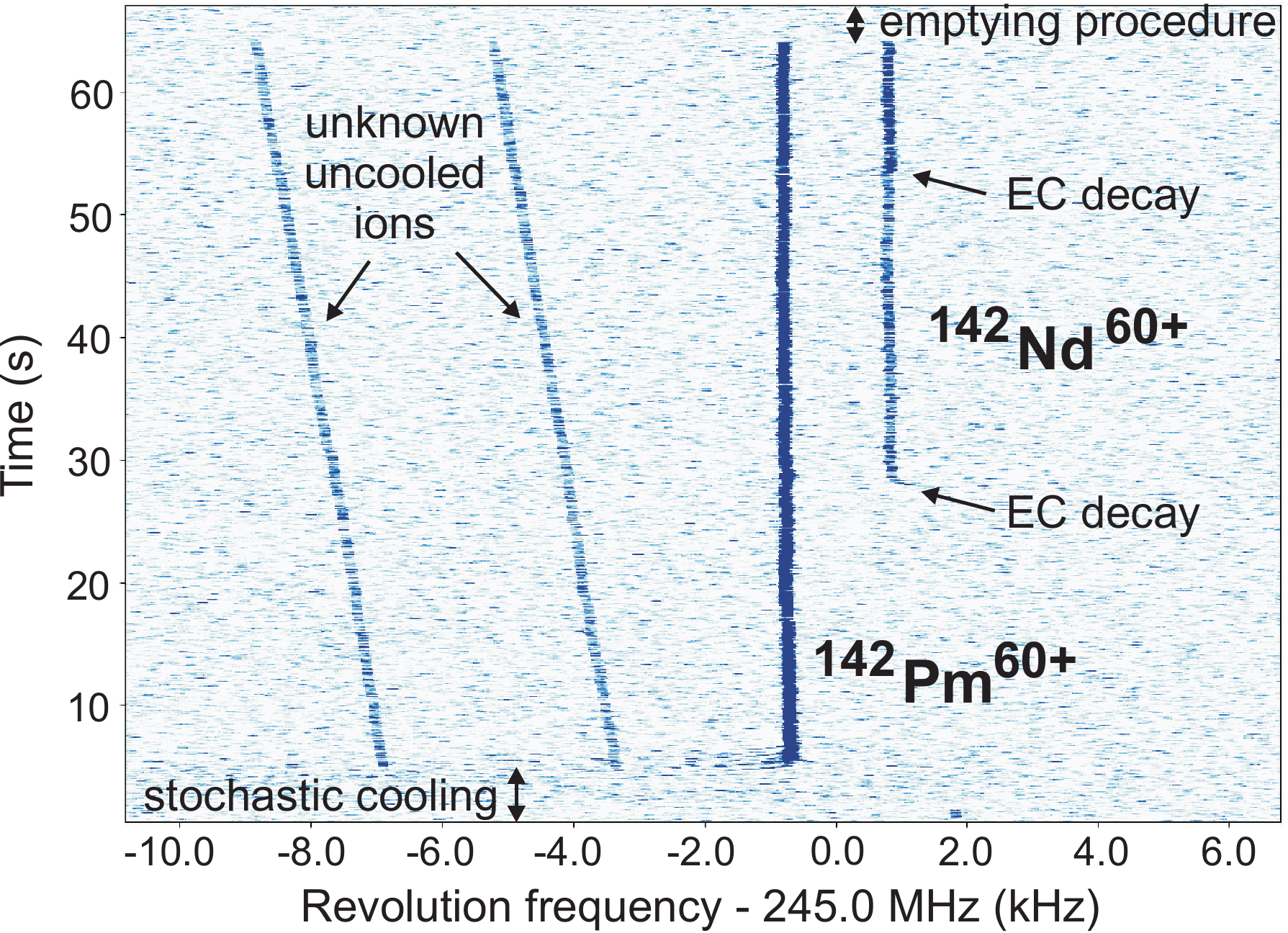}
\caption{Example of the measured traces of stored $^{142}$Pm$^{60+}$ and $^{142}$Nd$^{60+}$ ions in \red{the} ESR. 
The entire frequency bandwidth measured by RSA 5103A as well as the entire time range are shown.
The traces of $^{142}$Pm$^{60+}$ and $^{142}$Nd$^{60+}$ ions are clearly visible.
Two particles are observed which are uncooled (changing revolution frequency with time). 
The origin of these ions is not known. 
Since these particles are not cooled, their frequency does not correspond to their mass-to-charge ratio and thus their unambiguous identification is not possible.
\label{travellers}}
\end{figure*}

It is unlikely that the above arguments could influence the obtained results.
A possible effect of the larger number of stored particles has been checked by selecting files containing just 1 and 2 EC-decays. 
Although at inevitably lower counting statistics, no indications of a statistically significant modulation with $\omega\sim0.88$ rad s$^{-1}$ were found.

}
\section{Conclusion and Outlook}
In conclusion, the experiment repeated in 2014 does not confirm the observed $\approx20\%$ modulation amplitude in EC-decay of
hydrogen-like $\ce{^{142}Pm^{60+}}$ ions in \red{the} ESR reported in \cite{Litvinov2008}.
Furthermore, the new experiment confirms the results of the experiment in 2010, where the entire data set did not 
indicate the presence of a modulation \cite{Kienle2013}.


\redx{Our new Bayesian analysis of the older data does not confirm the statistical significances of the previously reported modulated decays.
This agrees with the recent findings \cite{bayes} that the $\chi^2$ minimisation results in much narrower confidence intervals and 
generates strong correlations between parameters as compared to the Bayesian approach.
On the one hand, the results of the new Bayesian analysis do not allow us to exclude the presence of the modulations in the older datasets.
On the other hand, the corresponding statistical significances are smaller than the ones obtained with the $\chi^2$ approach.
We also emphasise, that for data sets with large statistics, which are the full data set from 2010 and the one from this work, 
both methods do not indicate significant modulations.}

In the course of the experiments, data analysis, and interpretation of the results, numerous challenging cross-discipline problems 
have been realised and solved, which will be addressed in forthcoming publications.

If a new experiment would be possible in the future, it would be intriguing to investigate the second system studied in \cite{Litvinov2008}, namely $^{140}$Pr$^{58+}$.
The oscillation pattern was more clearly established in Pr as compared to Pm.
\redx{Another interesting system is the hydrogen-like $^{122}$I$^{52+}$, in which a modulation with a period of about 6 seconds might be present \cite{Faestermann2015,Kienle2009,Kienle2009b}.}
In addition to the storage ring complex of GSI, such an experiment might be possible at the storage ring CSRe in Lanzhou \cite{CSRe}.
Furthermore, experimental studies of weak decays are planned in the Electron Ion Beam Trap of the TITAN facility \cite{Leach2017}.

\section*{In Memoriam}

The authors will be ever grateful for the valuable contributions of
their late colleagues and friends Paul Kienle and Fritz Bosch, who enthusiastically
engaged in countless days of experiment preparation, shifts, data
analysis and discussion. They will always be remembered.

\section*{Acknowledgements}
We would like to thank the GSI accelerator and FRS teams for the excellent technical support and their steady help.
We are indebted to 
P.~Armbruster,
K.~Blaum, 
P.~Braun-Munzinger, 
C.~Ewerz,
M.~Faber, 
A.~F{\"a}{\ss}ler,
H.~Feldmeier, 
P.~Filip,
A.~Gal, 
F.~Giacosa, 
C.~Giunti,
W.~Greiner,
D.~von Harrach,
R.~Hayano,
W.~F.~Henning, 
A.~N.~Ivanov, 
V.~Ivanova,
H.~Lipkin, 
B.~Kayser,
H.~Kleinert, 
H.-J.~Kluge, 
J.~Kopp,
U.~K{\"o}ster,
R.~Kr{\"u}cken,
G.~Lambiase, 
K.~Langanke,
A.~Letourneau,
D.~Liesen,
M.~Lindner,
A.~Merle,
G.~M{\"u}nzenberg, 
W.~N{\"o}rtersh{\"a}user,
E.~W.~Otten,
Z.~Patyk, 
C.~Pe{\~n}a-Garay,
C.~Peschke, 
M.~Peshkin,
P.~Petri,
C.~Rappold, 
H.~Rauch, 
G.~Rempe, 
A.~Richter,
J.~P.~Schiffer,
H.~Schmidt, 
D.~Schwalm, 
V.~Soergel, 
H.~J.~Specht,
J.~Stachel,
B.~Stech,
H.~St{\"o}cker,
Y.~Tanaka,
N.~Troitskaya,
T.~Uesaka,
J.~Wambach,
Ch.~Weinheimer,
W.~Weise,
E.~Widmann,
G.~Wolschin,
T.~Yamazaki,
K.~Yazaki,
and
J.~Zmeskal
for fruitful discussions.
We thank ExtreMe Matter Institute EMMI of GSI for help in organising the dedicated EMMI Rapid Reaction Task Force.

This work was partially supported by 
the Scientific Research Project Coordination Unit of Istanbul University [Grant Numbers 48110, 54135, 53864 and FBA-2018-30033], 
the European Community FP7 Capacities: Research Infrastructures - Integrating Activity (IA) [Grant Number 262010 ``ENSAR''],
the DFG cluster of excellence ``Origin and Structure of the Universe'' of the Technische Universit{\"a}t M{\"u}nchen,
the Helmholtz-CAS Joint Research Group HCJRG-108,
\red{the Deutscher Akademischer Austauschdienst (DAAD) through Programm des Projektbezogenen Personenaustauschs (PPP) with China [Project ID 57389367],}
the External Cooperation Program of the Chinese Academy of Sciences [Grant Number GJHZ1305],
the HIC-for-FAIR through HGS-HIRE,
and Max Planck Society.
\redx{CB and CT acknowledge support by the German Federal Ministry of Education and Research (BMBF) [Grant Numbers 05P15RGFAA and 05P19RGFA1].}
TY acknowledges support by the Mitsubishi Foundation [Grant Number 23143], the Sumitomo Foundation [Grant Number 110501], and JSPS KAKENHI [Grant Numbers 26287036 and 17H01123].
YAL and RSS receive funding from the European Research Council (ERC) under the European Union's Horizon 2020 research and innovation programme 
[Grant Agreement Number 682841 ``ASTRUm''].
PMH acknowledges support by DFG [Project HI 2009/1-1].
CLW acknowledges support from the Austrian Science Fund \redx{[Grant Number J3503]}. 
PMW, PJW and CLW acknowledge support from the UK Science and Technology Facilities Council STFC. 
RBC acknowledges support from the Max-Planck Partner group.



 

 






\end{document}